\documentclass[journal]{IEEEtran}

\ifCLASSINFOpdf
\else
   \usepackage[dvips]{graphicx}
\fi
\usepackage{url}

\usepackage{microtype}
\usepackage{graphicx}
\usepackage{xcolor}
\usepackage{subfigure}
\usepackage{booktabs} 

\usepackage{hyperref}
\usepackage{float}

\usepackage{orcidlink}

\usepackage{amsmath}
\usepackage{amssymb}
\usepackage{amsfonts}
\usepackage{mathtools}
\usepackage{amsthm}
\usepackage{mathrsfs}
\usepackage{array}
\DeclareMathAlphabet{\mathpzc}{OT1}{pzc}{m}{it}

\usepackage[capitalize,noabbrev]{cleveref}

\usepackage{pifont}

\definecolor{darkgreen}{rgb}{0.0, 0.8, 0.0}
\definecolor{darkred}{rgb}{0.8, 0.0, 0.0}

\begin{document}

\title{Text-To-Speech with Chain-of-Details: modeling temporal dynamics in speech generation}

\author{Jianbo Ma\,\orcidlink{0000-0002-6765-9462}, Richard Cartwright\,\orcidlink{0009-0005-7627-0319}
   \thanks{Jianbo Ma is now with Canva research, work done in Dolby (e-mail: jianbo@canva.com).}
   \thanks{Richard Cartwright, work done with Dolby Laborotaries, He is now with Canva research. (e-mail: richardc@canva.com).
   }
}

\markboth{Journal of \LaTeX\ Class Files, Vol. 14, No. 8, August 2015}
{Shell \MakeLowercase{\textit{et al.}}: Bare Demo of IEEEtran.cls for IEEE Journals}
\maketitle

\begin{abstract}
Recent advances in Text-To-Speech (TTS) synthesis have seen the popularity of multi-stage approaches that first predict semantic tokens and then generate acoustic tokens. In this paper, we extend the coarse-to-fine generation paradigm to the temporal domain and introduce Chain-of-Details (CoD), a novel framework that explicitly models temporal coarse-to-fine dynamics in speech generation using a cascaded architecture. Our method progressively refines temporal details across multiple stages, with each stage targeting a specific temporal granularity. All temporal detail predictions are performed using a shared decoder, enabling efficient parameter utilization across different temporal resolutions. Notably, we observe that the lowest detail level naturally performs phonetic planning without the need for an explicit phoneme duration predictor. We evaluate our method on several datasets and compare it against several baselines. Experimental results show that CoD achieves competitive performance with significantly fewer parameters than existing approaches. Our findings demonstrate that explicit modeling of temporal dynamics with the CoD framework leads to more natural speech synthesis.
\end{abstract}


\IEEEpeerreviewmaketitle

\section{Introduction}
\label{sec:introduction}
\noindent Text-to-Speech (TTS) is a fundamental technology in modern Artificial Intelligence, powering a wide range of applications across diverse domains.

Over the past several decades, TTS research has evolved through multiple paradigms. Early efforts focused on articulatory synthesis~\cite{coker1976model}, followed by the development of statistical parametric synthesis systems~\cite{zen2009statistical}, such as those based on Hidden Markov Models (HMM)~\cite{tokuda2013speech, tokuda2000speech}. In the late 2010s, the introduction of neural vocoders that leverage architectures like SampleRNN~\cite{mehri2016samplernn} and WaveNet~\cite{van2016wavenet}, were employed by models like the Deep Voice series~\cite{arik2017deep, gibiansky2017deep, ping2018deep}, Tacotron series~\cite{wang2017tacotron, shen2018natural}, and Char2wav~\cite{sotelo2017char2wav}, which has improved quality of TTS. More recently, the powerful generative capabilities of diffusion models~\cite{sohl2015deep, ho2020denoising, song2020score} have been applied to TTS, as seen in works such as Glow-TTS~\cite{kim2020glow}, Grad-TTS~\cite{popov2021grad} and naturalSpeech 2~\cite{shen2024naturalspeech}.

The proliferation of large-scale speech datasets and advances in generative modeling have greatly enhanced the naturalness of synthesized speech. More recent state-of-the-art TTS systems can be broadly categorized into two groups: continuous feature generation and token prediction. Recent models like VoiceBox~\cite{le2023voicebox} and E2-TTS~\cite{eskimez2024e2}, which utilize flow-matching techniques, fall into the first category, while models such as VALL-E~\cite{wang2023neural}, NaturalSpeech 3~\cite{ju2024naturalspeech}, and MaskGCT~\cite{wang2024maskgct} represent the latter. Within the token prediction paradigm, approaches can be further divided into autoregressive (AR) and non-autoregressive (NAR) models. AR models, which generate tokens sequentially in a manner akin to large language models (LLMs), have attracted considerable attention; Parler-TTS~\cite{lyth2024natural} is a notable example. In contrast, NAR models offer the advantage of parallel decoding, resulting in faster inference. Notably, SPEAR-TTS~\cite{kharitonov2023speak} introduced a two-stage, coarse-to-fine generation process, where the first stage predicts semantic information and the second stage refines acoustic details. This multi-stage paradigm has been adopted by subsequent works such as NaturalSpeech 3~\cite{ju2024naturalspeech} and MaskGCT~\cite{wang2024maskgct}. Additionally,~\cite{gallego2025single} proposed a single-stage model employing Masked Token Modeling inspired by MaskGIT~\cite{chang2022maskgit}, though some trade-offs in performance has been observed.

In this paper, we extend the coarse-to-fine generation paradigm by explicitly incorporating temporal dynamics through the introduction of the Chain-of-Detail (CoD) method. Inspired by VAR~\cite{tian2024visual}, which models the progressive increase of image resolution in an autoregressive manner, our CoD approach decomposes speech generation into multiple progressive stages. At each stage, the model generates a version of the speech representation at a particular temporal resolution—starting from a temporal coarse, low-detail representation that captures broad timing and structure, and progressively refining it to higher temporal resolutions that add finer-grained acoustic details. Unlike VAR in the image domain, each temporal level in CoD employs masked generative modeling. 
Crucially, in CoD, a unified codebook and a single model architecture can be shared to all stages. This design allows the model to efficiently reuse parameters across all temporal resolutions, improving both efficiency and consistency. During inference, CoD-TTS operates in a non-autoregressive manner, generating all tokens in parallel at each temporal level. Experimental results show that CoD-TTS achieves competitive performance with fewer parameters.

\section{Related Works}
\label{sec:related_works}

\begin{figure}[!t]
    \centering
    \includegraphics[width=0.8\linewidth]{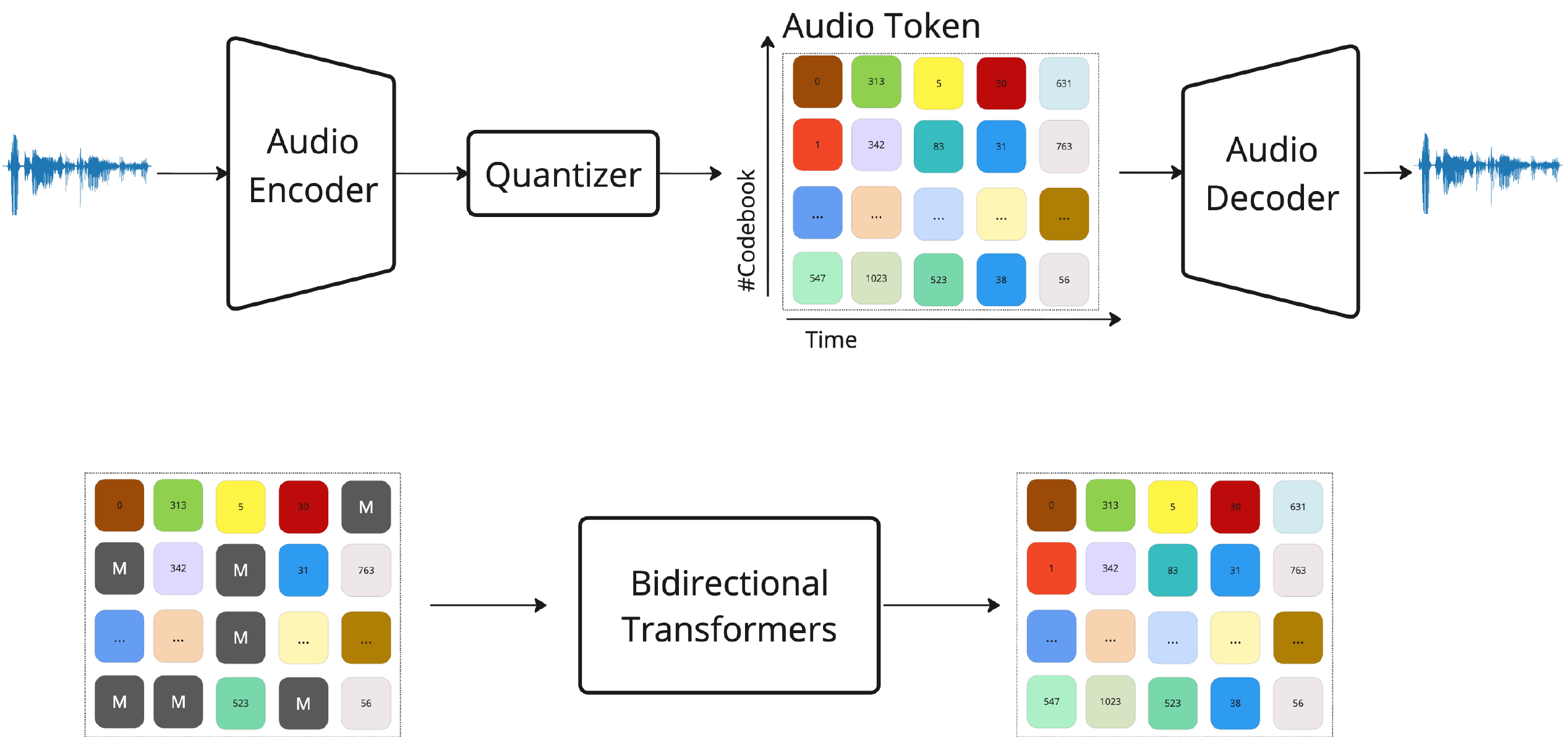}
    \caption{Pipeline Overview of masked audio token modeling approach. It consists of two stages with 1) tokenization that tokenizes audio into audio tokens, and 2) masked audio token modeling predicts audio tokens masked at random with a transformer decoder.}
    \label{fig:masked_audio_token_modeling}
\end{figure}

\noindent As the proposed CoD model is a discrete token-based model, we primarily focus on related work in this category. Modern TTS models typically involve two main stages: speech representation and generative modeling. The first stage employs a VQ-GAN architecture where an encoder maps continuous time-domain waveforms into lower-dimensional representation embeddings, followed by vector quantization using techniques such as Residual Vector Quantization (RVQ)~\cite{zeghidour2021soundstream}. The quantized representations are subsequently converted back to continuous time-domain waveforms by a decoder. Representative works in this area include SoundStream~\cite{zeghidour2021soundstream}, EnCodec~\cite{defossez2022high}, and DAC~\cite{kumar2023high}. Figure~\ref{fig:masked_audio_token_modeling} upper figure shows this first stage.

The subsequent generative models operate on the discrete token space. Parler-TTS, which is a public implementation based on~\cite{lyth2024natural}, is an autoregressive token prediction model that utilizes the delay pattern introduced in MusicGen~\cite{copet2023simple}. It follows a decoder-only architecture similar to~\cite{brown2020language}. We refer to this type of model that directly predicts acoustic tokens from transcript conditions as one-stage models.

\noindent \textbf{Masked Audio Token Modeling:} Inspired by MaskGIT~\cite{chang2022maskgit}, Masked Audio Token Modeling (MATM) learns to predict audio tokens that are randomly masked during training. Figure~\ref{fig:masked_audio_token_modeling} presents the pipeline overview of this approach. During training, random audio tokens are masked and fed into a bidirectional transformer decoder, which learns to predict the masked tokens using cross-entropy loss. During inference, a cosine masking schedule is employed to iteratively predict audio tokens across multiple steps, where each step samples predicted tokens based on their confidence probabilities. We encourage reader to refer~\cite{chang2022maskgit} for details.

\cite{gallego2025single} proposed a single-stage masked audio token modeling approach that leverages knowledge distillation from a two-stage teacher model. This single-stage framework shares structural similarities with each level of our proposed CoD method, as both employ masked token prediction for speech generation. However, the temporal modeling aspect is not explicitly addressed in this work.

\noindent \textbf{Multi-stage generative models:} The concept of two-stage TTS models was first introduced in SPEAR-TTS~\cite{kharitonov2023speak}. The core idea involves first predicting semantic tokens from text, followed by generating acoustic tokens conditioned on these semantic representations. This approach draws inspiration from AudioLM~\cite{borsos2023audiolm}, which proposed generating acoustic tokens through a coarse-to-fine progression from semantic tokens to detailed acoustic representations using a three-stage design. While originally developed for autoregressive (AR) language model-based generative models, this coarse-to-fine approach has been successfully adapted to non-autoregressive (NAR) models as well. 

Several language model-based TTS systems employ multiple cascaded stages. Examples include VALL-E~\cite{wang2023neural}, MaskGCT~\cite{wang2024maskgct}, and NaturalSpeech 3~\cite{ju2024naturalspeech}. VALL-E~\cite{wang2023neural} first uses an AR language model to predict the acoustic token sequence of the first layer of RVQ, followed by NAR modeling to predict the remaining acoustic token sequences corresponding to the other RVQ layers. Similarly, MaskGCT~\cite{wang2024maskgct} employs NAR modeling to first predict semantic tokens, then uses a second model to predict the first layer of RVQ, and finally applies a third model to predict the remaining acoustic token sequences. NaturalSpeech 3~\cite{ju2024naturalspeech} disentangles speech into different token sequences that correspond to different speech attributes, with different models cascaded to predict each token sequence.

Although multi-stage NAR models share conceptual similarities with our proposed CoD method in terms of progressive generation, they do not explicitly model temporal coarse-to-fine dynamics during both training and inference. In previous methods, the 'coarse' token refers to a subset of residual quantizer layers in the RVQ of the audio neural codec and does not encompass temporal aspects. Unlike existing approaches that primarily focus on semantic-to-acoustic token conversion, our CoD method specifically designs the modeling process to capture and leverage the inherent temporal dynamics of speech generation.

\begin{figure*}[!ht]
   \centering
   \includegraphics[width=\textwidth]{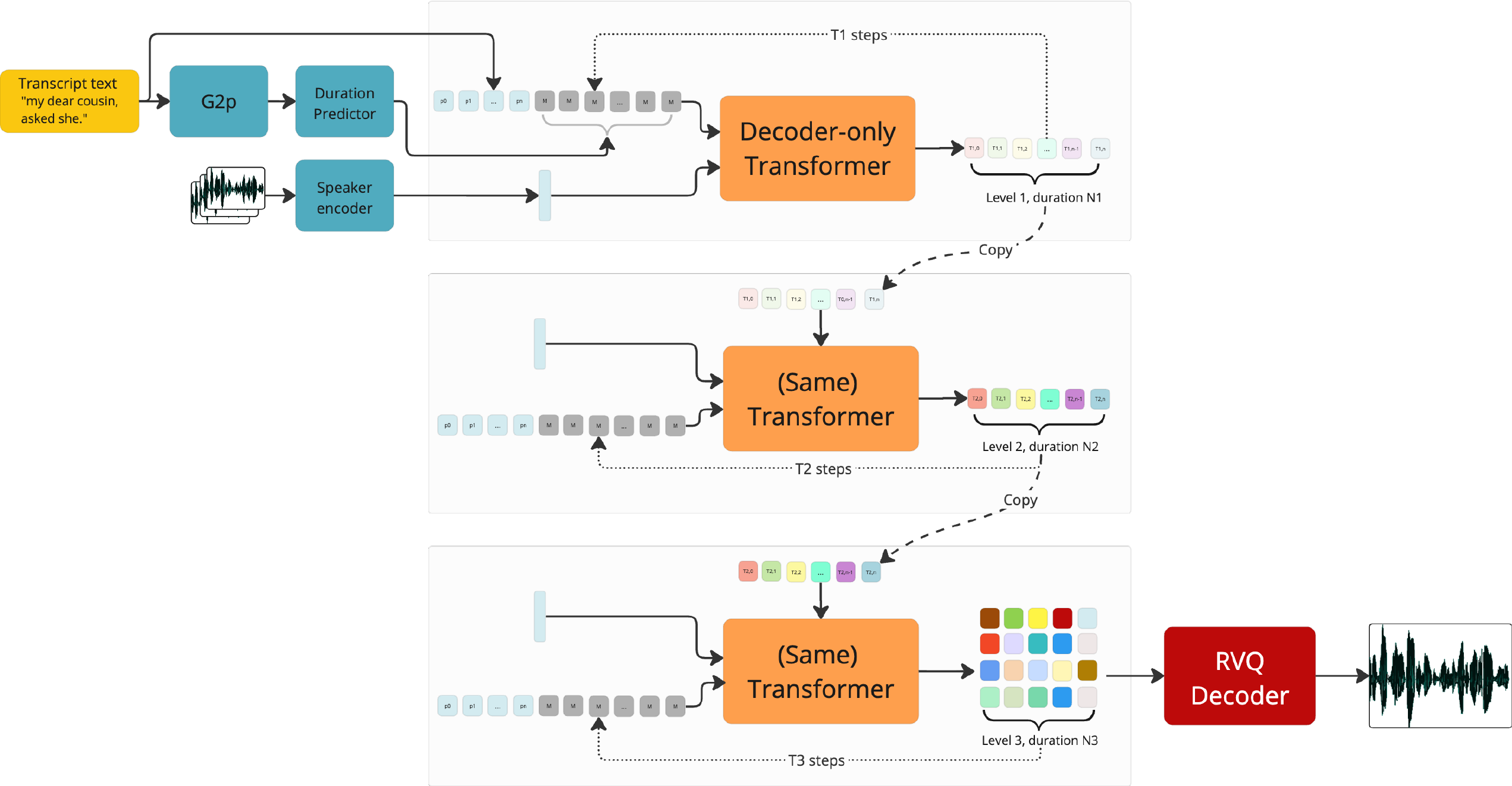}
   \caption{Overview of the Chain-of-Details (CoD) TTS inference pipeline with three temporal levels. During inference, a pretrained grapheme-to-phoneme (G2P) model generates phonetic indices from the transcript, and a pretrained duration predictor estimates the utterance length. Speaker embedding, extracted from a pretrained speaker encoder, conditions the system on the desired speaker identity. At temporal level 1, the transformer decoder interatively predicts masked audio tokens based on the G2P and speaker information, operating at token rate $f_{1}$ for $T_{1}$ steps. At temporal level 2, the same decoder refines the output by predicting masked audio tokens conditioned on both the previous level's output and the same conditioning information, at token rate $f_{2}$ for $T_{2}$ steps. This process continues for subsequent temporal levels, progressively increasing temporal resolution and detail in the generated speech.}
   \label{fig:CoD}
\end{figure*}

\section{Methods}

\subsection{Model overview}
\noindent The proposed Chain-of-Details (CoD) structure is show in Figure~\ref{fig:CoD}. Overall, it still follows 2-stages modeling as show in Figure~\ref{fig:masked_audio_token_modeling} that first convert speech into token space and then followed by the Masked Audio Token Modeling (MATM) stage. Unlike previous method, the discrete token prediction stage employs a cascaded architecture that progressively refines temporal details from coarse to fine resolution, where each stage focuses on a specific temporal granularity. 

\subsection{Chain-of-Details: tempoal coarse-to-finer modeling}
\label{sec: CoD}
We draw inspiration from artists creating drawings, which typically begin with an outline before gradually adding finer details. Unlike VAR~\cite{tian2024visual} that progressively increase resolution in image generation and at each resolution stage, there is no iterative prediction, CoD involves breaking down speech generation into several progressive stages, each increasing the temporal resolution of the generated speech and iteratively predicting masked audio tokens. The proposed CoD combines the strengths of both MATM and temporal dynamics modeling. As illustrated in Figure~\ref{fig:CoD}, we take three temporal resolutions as an example, where we refer them as temporal level 1, temporal level 2 and temporal level 3, with higher index representing finer temporal resolution. 

Let $X_l = [x^{l}_1, x^{l}_2, \ldots, x^{l}_{t_l}]$ denote the sequence of audio tokens for an utterance at temporal level $l$, where $N_l$ is the token sequence length at that level. We define the temporal token rate as $f_l$, representing the number of tokens per second. As temporal resolution increases with higher levels, it follows that $f_1 < f_2$ and $N_1 < N_2$. During training, a subset of the target audio tokens is selected and replaced with a special [MASK] token if a sampled masked $m_i=1$; otherwise, the original token is retained. Following the approach in~\cite{chang2022maskgit}, the mask generation is parameterized by a schedule function such as the cosine schedule function. After masking, we denote the resulting sequence as $X^{'}_l$. The model is then trained to minimize the negative log-likelihood of the masked tokens with model parameter $\theta$:


\begin{equation}
    \label{eq:masked_loss2}
    \begingroup\makeatletter\def\f@size{6}\check@mathfonts
    \def\maketag@@@#1{\hbox{\m@th\large\normalfont#1}}%
    \begin{aligned}
        \scriptstyle \mathcal{L}_{mask} = \left \{ \begin{array}{rcl} 
            -\mathbb{E}_{\mathbf{X_{l} \in \mathpzc{D}}} [ \sum_{\forall i \in [0, t_{l}-1], m_{i}=1} \log p_{\theta}(x^{l}_i | X^{'}_l, X_{l-1}, C ) ] & \textnormal{for} \quad l>1, \\[1.5ex] 
            -\mathbb{E}_{\mathbf{X_{l} \in \mathpzc{D}}} [ \sum_{\forall i \in [0, t_{l}-1], m_{i}=1} \log p_{\theta}(x^{l}_i | X^{'}_l, C ) ] & \textnormal{for} \quad l=1, 
        \end{array} \right.
    \end{aligned}
    \endgroup
\end{equation}
where $C$ denotes other conditions such as transcript.

During training, the temporal level $l$ is randomly selected, with a bias towards higher temporal resolutions. To further enhance robustness of token prediction, we augment the conditioning audio tokens from the previous temporal level ($X_{l-1}$ for $l > 1$) by randomly replacing a portion of these tokens with randomly chosen indices within the vocabulary.

As illustrated in Figure~\ref{fig:CoD}, the inference process begins by utilizing a pretrained grapheme-to-phoneme (G2P) model to convert the transcript into phonetic indices, a pretrained duration predictor to estimate the utterance length, and a pretrained speaker encoder to extract the speaker embedding. At temporal level 1, the phonetic indices are concatenated with a sequence of fully masked audio tokens, whose length is determined by the estimated duration. All conditioning information is provided to the transformer decoder, to generates $p_{\theta}(x^{1}_i | X^{'}_{1,t_{l}}, C )$, where $X^{'}_{1,t_{l}}$ consists entirely of [MASK] tokens at $t_{l}=1$. Token indices are sampled, and their associated probabilities serve as confidence scores. Based on the mask schedule function, the number of masked tokens for $t_{l}=2$ is determined, and tokens with the highest confidence scores are selected and fixed to form $X^{'}_{1,2}$. This iterative process is repeated until $t_{1}=T_1$. 

At temporal level 2, the same decoder further refines the output by predicting masked audio tokens conditioned on both the output from the previous level and the same set of conditioning information, operating at token rate $f_{2}$ for $T_{2}$ steps. This can be expressed as $p_{\theta}(x^{2}_i | X^{'}_{2,t_{2}}, X_{1}, C )$. The process is repeated for subsequent temporal levels, progressively increasing the temporal resolution and detail of the generated speech.

\begin{table*}[!ht]
   \centering
   \caption{Experimental results on LibriSpeech test-clean.}
   \small
   \def\arraystretch{1.0}
   \setlength\tabcolsep{1 pt}
   \scalebox{1.0}{
       \begin{tabular}{lccccc}
           \toprule
           \textbf{Name}
           & \textbf{Params. (M)} 
           & \textbf{WER}($\downarrow$)(4-10s)
           & \textbf{Training Data}
           & \textbf{Hours}\\
           \midrule
           Ground truth
           & -  
           & 2.2 
           & -
           & -
           \\
           DAC recon.
           & -
           & 2.4  
           & - 
           & -
           \\
           \midrule
           YourTTS
           & -
           & 7.1  
           & LibriTTS-clean+Pt+Fr 
           & 474
           \\
           VALL-E
           & 370
           & 5.9  
           & LibriLight 
           & 60k
           \\
           StyleTTS 2
           & -
           & 4.0  
           & LibriTTS-clean 
           & 245
           \\
           KD-NARSIS
           & 249
           & 5.9   
           & LibriTTS-clean 
           & 245
           \\
           NAR 2-stage
           & 476
           & 3.6   
           & LibriTTS-clean 
           & 245
           \\
           \midrule
           \textbf{CoD-Base} (Ours)
           & 263
           & 3.09
           & LibriTTS-clean
           & 245
           \\
           \textbf{CoD-Large} (Ours)
           & 503
           & 2.81
           & LibriTTS-clean + MLS-En-Clean
           & 3297
           \\
           \bottomrule
       \end{tabular}
    }
    \label{tab:exp_results_librispeech}
\end{table*}

\section{Experiments}
\label{sec:exp}

\subsection{Models}
\label{sec:method_models}

\noindent \textbf{Tokenizer} As discussed in Section~\ref{sec: CoD}, our approach requires an audio tokenizer capable of representing temporal structure. Specifically, the tokenizer must encode audio into $L$ temporal levels, each with its own token rate $f_l$. We explored two strategies to achieve this.

The first method leverages the Residual Vector Quantization (RVQ) structure. In RVQ, higher-layer quantizers encode the residuals of lower layers, so tokens from the lower quantizers capture more information. We adopted the DAC~\cite{kumar2023high} model with 9 codebooks, operating at 8 kbps, a 44.1 kHz sampling rate, and a token rate of $86.13$ Hz. To obtain lower temporal resolution tokens for CoD, we decimated the token sequence from the first quantizer of the RVQ. For instance, in the $L=3$ setting, for temporal level $l=2$, we downsampled the first-layer token sequence by a factor of 2, resulting in a token rate $f_2=43.07$ Hz. For temporal level $l=1$, we used a decimation factor of 4, yielding $f_1=21.53$ Hz.

Additionally, we designed a tokenizer that explicitly models temporal hierarchy in audio. This approach adds $L-1$ extra quantizers to the original RVQ in DAC. For example, in the $L=3$ setting, the quantizer for $l=2$ encodes the audio representation of $l=3$ after a decimation layer, and the quantizer for $l=1$ encodes the representation of $l=2$ after further decimation. The codebooks for quantizers at different temporal levels can either be shared or independent. Experimental results comparing these tokenization strategies are presented in Table~\ref{tab:ablation_token_types}.

\noindent \textbf{G2P} We used a pre-trained SoundChoice G2P~\cite{ploujnikov2022soundchoice} from Speechbrain~\cite{speechbrain}.

\noindent \textbf{Duration predictor} The duration predictor utilizes phonetic information from the G2P module to estimate durations in seconds. Its architecture is intentionally lightweight, following the design of~\cite{gallego2024single}, and consists of only 6-layers transformer layers with hidden dimension as 256.

\noindent \textbf{Speaker encoder} We used the pretrained Wespeaker~\cite{wang2023wespeaker} as speaker embedding extractor.

\noindent \textbf{Transformer decoder} We use a Llama-style~\cite{touvron2023llama, lyth2024natural} Transformer architecture as the backbone, but replacing causal attention with bidirectional attention. Inspired from DiT~\cite{peebles2023scalable}, we use adaptive LayerNorm~\cite{perez2018film} for speaker embedding condition.

\subsection{Implementation details}

\noindent \textbf{Training details} We use a batch size of 256 with a learning rate of 1e-4 and a cosine scheduler with 4000 warm-up steps. The AdamW optimizer~\cite{loshchilov2017decoupled} with $\beta_1 = 0.9$, $\beta_2 = 0.95$, and weight decay as $0.05$ is employed for all experiments. All models are trained for 400K steps. During training, we randomly sample different temporal coarse levels with biased probabilities to ensure balanced training across all levels. For example, for the three-level temporal coarse hierarchy, we use sampling probabilities of [0.2, 0.3, 0.5] for the coarsest to finest levels respectively, allowing simultaneous training of all temporal coarse levels. Classifier-Free Guidance (CFG)~\cite{ho2022classifier} is also employed, with a 10\% conditioning dropout, where transcript and previous coarse level conditionings are replaced by a learnable embeddings. As illustrated in Section~\ref{sec: CoD}, we conduct augmentation of the conditioning audio tokens from the previous temporal level by 10\%.

\noindent \textbf{Inference details} The inference procedue follows ~\cite{gallego2024single} and also based on based on the sampling method of MaskGIT~\cite{chang2022maskgit}. Each level we utilized 20 steps. CFG is applied during inference. But we decrease the guidance level linearly from 3 to 0.75 from start decoding to the end. To add diversity to the output speech, Gaussian noise with zero mean and variance linearly decreasing from 3.0 to 0 is applied to logits during decoding.

\subsection{Data}

\noindent For our experiments, we utilized the LibriTTS~\cite{zen2019libritts} dataset and a subset of the MLS~\cite{pratap2020mls} English corpus. To ensure high data quality, we applied filtering criteria, selecting only audio samples from the MLS English subset with a signal-to-noise ratio (SNR) greater than 55dB and C50 larger than 55. Followed~\cite{lyth2024natural} we used the Brouhaha library~\cite{lavechin2023brouhaha} to estmate those values. This resulted in approximately 3,000 hours of high-quality speech data. All audio was obtained at a 44.1 kHz sampling rate as full-band speech from the LibriVox\footnote{https://librivox.org/}, ensuring consistency in audio quality and format throughout our training corpus.

We trained two model variants: a base model and a large model. The base model consists of 12 transformer layers in the generative transformer decoder, each with a hidden dimension of 1024 and is trained with LibriTTS-clean only. The large model employs 24 transformer layers and is trained with both LibriTTS-clean and MLS English clean subset. Further details of the experimental setup are provided in Table~\ref{tab:exp_setup}.

\begin{table*}[t]
    \centering
    \caption{Experimental setup.}
    \small
    \def\arraystretch{1.0}
    \setlength\tabcolsep{1 pt}
    \scalebox{1.0}{
    \begin{tabular}{lcccccc}
        \toprule
        \textbf{Model}
            & \textbf{Params. (M)} 
            & \textbf{Training Data}
            & \textbf{Hours}
            & \textbf{Sampling Rate (KHz)}
            & \textbf{Ph. Tokenizer}
            & \textbf{BPE Tokenizer}\\
        \midrule
        Base
        & 263  
        & LibriTTS-clean 
        & 245 
        & 44.1
        & \textcolor{darkgreen}{\ding{51}}
        & \textcolor{darkgreen}{\ding{51}}
        \\
        Large
        & 502 
        & LibriTTS-clean+MLS\_En-clean  
        & 3297
        & 44.1
        & \textcolor{darkred}{\ding{55}}
        & \textcolor{darkgreen}{\ding{51}}
        \\
        \bottomrule
    \end{tabular}}
    \label{tab:exp_setup}
\end{table*}

\textbf{Evaluation data:} In line with previous works~\cite{borsos2023audiolm, wang2023neural, defossez2022high}, we primarily evaluate our models on samples from the LibriSpeech test-clean set~\cite{panayotov2015librispeech} with durations between 4 and 10 seconds and compare with baseline systems. Additionally, we assess performance on the SeedTTS test set~\cite{anastassiou2024seed} and conduct ablation studies on the test-clean subset of LibriTTS.

\subsection{Evaluation metrics}
We follow the~\cite{gallego2025single} to evaluate the effectiveness of the proposed CoD-TTS.

\textbf{WER} We use Word Error Rate to evaluate the intelligibility of generated speech.

\textbf{MOS} We conducting human evaluation the collect Mean Opinion Score.

\subsection{Experimental results on LibriSpeech}

Table~\ref{tab:exp_results_librispeech} presents the experimental results comparing our proposed CoD-TTS models with existing baseline methods on LibriSpeech test-clean dataset. Our CoD-Base model achieves a WER of 3.09\% with 263M parameters, significantly outperforming comparable methods like KD-NARSIS (5.9\% WER with 249M parameters) and StyleTTS 2 (4.0\% WER) when trained on the same LibriTTS-clean dataset. The CoD-Large model further improves performance to 2.81\% WER with 503M parameters, approaching the quality of ground truth (2.2\% WER) and DAC reconstruction (2.4\% WER). Notably, our method demonstrates superior parameter efficiency compared to VALL-E, achieving better WER with fewer parameters and significantly less training data (245 hours vs 60k hours). When compared with a baseline with 2 stage model using the same training data (NAR 2-stage), our method achieved better WER with nearly half number of parameters. Compared with the knowledge distillation 1-stage method (KD-NARSIS), with same level of parameter number, the WER has been observed dropped significantly.

\subsection{Experimental results on SeedTTS test-set}
\label{sec:exp_results}

\begin{table}[h]
    \centering
    \caption{Experimental results on SeedTTS test-set.}
    \small
    \def\arraystretch{1.0}
    \setlength\tabcolsep{1 pt}
    \scalebox{0.75}{
        \begin{tabular}{lccc}
            \toprule
            \textbf{Name}
                & \textbf{Params. (M)} 
                & \textbf{WER} ($\downarrow$)
                & \textbf{Training Data Hours}
                \\
            \midrule
            Ground truth
            & -  
            & 2.14 
            & -
            \\
            DAC recon.
            & -
            & 2.60 
            & - 
            \\
            \midrule
            MaskGCT
            & 1B
            & 2.62
            & Emilia (100k)
            \\
            \midrule
            \textbf{CoD-Base} (Ours)
            & 263
            & 2.89
            & LibriTTS-clean (245)
            \\
            \textbf{CoD-Large} (Ours)
            & 503
            & 2.73
            & LibriTTS-clean+MLS-En-Clean (3297)
            \\
            \bottomrule
        \end{tabular}
    }
    \label{tab:exp_results_seedtts}
\end{table}

Table~\ref{tab:exp_results_seedtts} presents the experimental results on the SeedTTS test-set~\cite{anastassiou2024seed}, which provides an additional evaluation benchmark for our proposed CoD-TTS models. Our CoD-Base model achieves a WER of 2.89\%, demonstrating competitive performance with significantly fewer parameters (263M) compared to MaskGCT (1B parameters). Note that MaskGCT employs a three-stage architecture: the first stage predicts semantic tokens, the second stage predicts the first layer of RVQ acoustic tokens, and the third stage predicts the remaining RVQ acoustic tokens. Despite this multi-stage complexity, our CoD approach achieves comparable results with substantially fewer parameters. The CoD-Large model further improves performance to 2.73\% WER while using significantly less training data (3,297 hours vs 100k hours for MaskGCT). These results confirm the effectiveness and parameter efficiency of our Chain-of-Details approach across different evaluation datasets.

\subsection{Ablation study}
In order to study the effects of different temporal coarse levels and the type of tokens, we have conducted two ablation studies. Note that the results are evaluated on the LibriTTS test-clean set. We used the CoD-Base model for all ablation study.

We have conducted an ablation study on the number of levels in CoD structure. The results are show in Table~\ref{tab:ablation_num_levels}. 

\begin{table}[H]
    \centering
    \caption{Ablation with number of levels.}
    \small
    \def\arraystretch{1.0}
    \setlength\tabcolsep{1 pt}
    \scalebox{0.75}{
    \begin{tabular}{lcc}
        \toprule
        \textbf{Name}
            & \textbf{WER}($\downarrow$)(4-10s)
            & \textbf{WER}($\downarrow$)(all)
            \\
        \midrule
        Decimated-with-3-levels
        & 3.78
        & 4.88
        \\
        Decimated-with-2-levels
        & 4.00
        & 5.19
        \\
        1-levels
        & 4.64
        & 7.67
        \\
        \bottomrule
    \end{tabular}}
    \label{tab:ablation_num_levels}
\end{table}

Here we used the decimated first layer of RVQ acoustic token as the temporal coarse token. We studied the 1 level, 2 levels and 3 levels. We did not go beyond 4 levels mainly because for most of utterances, the length of the lowest layer tokens to be predicted is smaller than the number of phonemes. From Table~\ref{tab:ablation_num_levels}, we observed that more levels tend to achieve better performance with large margin from 1 level to 2 levels.

\begin{table}[!h]
    \centering
    \caption{Ablation with coarse token types.}
    \small
    \def\arraystretch{1.0}
    \setlength\tabcolsep{1 pt}
    \scalebox{0.75}{
    \begin{tabular}{lcc}
        \toprule
        \textbf{Name}
            & \textbf{WER}($\downarrow$)(4-10s)
            & \textbf{WER}($\downarrow$)(all)
            \\
        \midrule
        Decimated-with-3-levels
        & 3.78
        & 4.88
        \\
        C2C1-Ind-trained
        & 5.81
        & 7.03
        \\
        C2C1-ShareCodebook
        & 7.99
        & 8.48
        \\
        HuBERT-2-levels
        & 4.62
        & 5.61
        \\
        \bottomrule
    \end{tabular}}
    \label{tab:ablation_token_types}
\end{table}

Table~\ref{tab:ablation_token_types} shows the results on different types of temporal coarse tokens. Here we studied the decimated first layer of RVQ acoustic token with tempral decimation factor of 2, independently trained level tokens that has different codebooks, temporal coarse tokens that share the same codebook, and HuBERT token with 1 additional coarse temporal level. It has been seen that the directly decimated acoustic token achieved best performance. It also shows that HuBERT token as temporal coarse token also achieves comparable results. But the independently trained temporal coarse tokens perform worse than the rest two. We hypothesis that the inferior performances of independently trained temporal coarse may tokens may caused by the fact that it lacks of direct mapping relationship with the final acoustic tokens or it has not been converged well. We opt to investigate it in future work.

\section{Conclusion}

\noindent This paper presented Chain-of-Details (CoD), a novel framework that extends the coarse-to-fine generation paradigm into the temporal domain for Text-to-Speech synthesis. Our experimental results demonstrate the effectiveness of CoD: the CoD-Base model surpasses comparable single-stage and two-stage methods, while also achieving superior parameter efficiency. Ablation studies further confirm the benefits of modeling with multiple temporal levels. For future work, we plan to investigate the optimal number of temporal levels, to develop improved temporal tokenization strategies, and expand the CoD framework to other modalities such as video generation.








\bibliographystyle{IEEEtran}
\bibliography{cod_tts}

@article{le2023voicebox,
  title={Voicebox: Text-guided multilingual universal speech generation at scale},
  author={Le, Matthew and Vyas, Apoorv and Shi, Bowen and Karrer, Brian and Sari, Leda and Moritz, Rashel and Williamson, Mary and Manohar, Vimal and Adi, Yossi and Mahadeokar, Jay and others},
  journal={Advances in neural information processing systems},
  volume={36},
  pages={14005--14034},
  year={2023}
}

@inproceedings{eskimez2024e2,
  title={E2 tts: Embarrassingly easy fully non-autoregressive zero-shot tts},
  author={Eskimez, Sefik Emre and Wang, Xiaofei and Thakker, Manthan and Li, Canrun and Tsai, Chung-Hsien and Xiao, Zhen and Yang, Hemin and Zhu, Zirun and Tang, Min and Tan, Xu and others},
  booktitle={2024 IEEE Spoken Language Technology Workshop (SLT)},
  pages={682--689},
  year={2024},
  organization={IEEE}
}

@article{kharitonov2023speak,
  title={Speak, read and prompt: High-fidelity text-to-speech with minimal supervision},
  author={Kharitonov, Eugene and Vincent, Damien and Borsos, Zal{\'a}n and Marinier, Rapha{\"e}l and Girgin, Sertan and Pietquin, Olivier and Sharifi, Matt and Tagliasacchi, Marco and Zeghidour, Neil},
  journal={Transactions of the Association for Computational Linguistics},
  volume={11},
  pages={1703--1718},
  year={2023},
  publisher={MIT Press One Broadway, 12th Floor, Cambridge, Massachusetts 02142, USA~…}
}

@article{borsos2023audiolm,
  title={Audiolm: a language modeling approach to audio generation},
  author={Borsos, Zal{\'a}n and Marinier, Rapha{\"e}l and Vincent, Damien and Kharitonov, Eugene and Pietquin, Olivier and Sharifi, Matt and Roblek, Dominik and Teboul, Olivier and Grangier, David and Tagliasacchi, Marco and others},
  journal={IEEE/ACM transactions on audio, speech, and language processing},
  volume={31},
  pages={2523--2533},
  year={2023},
  publisher={IEEE}
}

@article{wang2024maskgct,
  title={MaskGCT: Zero-shot text-to-speech with masked generative codec transformer},
  author={Wang, Yuancheng and Zhan, Haoyue and Liu, Liwei and Zeng, Ruihong and Guo, Haotian and Zheng, Jiachen and Zhang, Qiang and Zhang, Xueyao and Zhang, Shunsi and Wu, Zhizheng},
  journal={arXiv preprint arXiv:2409.00750},
  year={2024}
}

@article{gallego2024single,
  title={Single-stage TTS with Masked Audio Token Modeling and Semantic Knowledge Distillation},
  author={G{\'a}llego, Gerard I and Fejgin, Roy and Yeh, Chunghsin and Liu, Xiaoyu and Bhattacharya, Gautam},
  journal={arXiv preprint arXiv:2409.11003},
  year={2024}
}

@article{ju2024naturalspeech,
  title={Naturalspeech 3: Zero-shot speech synthesis with factorized codec and diffusion models},
  author={Ju, Zeqian and Wang, Yuancheng and Shen, Kai and Tan, Xu and Xin, Detai and Yang, Dongchao and Liu, Yanqing and Leng, Yichong and Song, Kaitao and Tang, Siliang and others},
  journal={arXiv preprint arXiv:2403.03100},
  year={2024}
}

@article{tian2024visual,
  title={Visual autoregressive modeling: Scalable image generation via next-scale prediction},
  author={Tian, Keyu and Jiang, Yi and Yuan, Zehuan and Peng, Bingyue and Wang, Liwei},
  journal={arXiv preprint arXiv:2404.02905},
  year={2024}
}

@article{wang2023neural,
  title={Neural codec language models are zero-shot text to speech synthesizers},
  author={Wang, Chengyi and Chen, Sanyuan and Wu, Yu and Zhang, Ziqiang and Zhou, Long and Liu, Shujie and Chen, Zhuo and Liu, Yanqing and Wang, Huaming and Li, Jinyu and others},
  journal={arXiv preprint arXiv:2301.02111},
  year={2023}
}

@article{lyth2024natural,
  title={Natural language guidance of high-fidelity text-to-speech with synthetic annotations},
  author={Lyth, Dan and King, Simon},
  journal={arXiv preprint arXiv:2402.01912},
  year={2024}
}

@inproceedings{gallego2025single,
  title={Single-stage TTS with Masked Audio Token Modeling and Semantic Knowledge Distillation},
  author={G{\'a}llego, Gerard I and Fejgin, Roy and Yeh, Chunghsin and Liu, Xiaoyu and Bhattacharya, Gautam},
  booktitle={ICASSP 2025-2025 IEEE International Conference on Acoustics, Speech and Signal Processing (ICASSP)},
  pages={1--5},
  year={2025},
  organization={IEEE}
}

@inproceedings{chang2022maskgit,
  title={Maskgit: Masked generative image transformer},
  author={Chang, Huiwen and Zhang, Han and Jiang, Lu and Liu, Ce and Freeman, William T},
  booktitle={Proceedings of the IEEE/CVF conference on computer vision and pattern recognition},
  pages={11315--11325},
  year={2022}
}

@article{zeghidour2021soundstream,
  title={Soundstream: An end-to-end neural audio codec},
  author={Zeghidour, Neil and Luebs, Alejandro and Omran, Ahmed and Skoglund, Jan and Tagliasacchi, Marco},
  journal={IEEE/ACM Transactions on Audio, Speech, and Language Processing},
  volume={30},
  pages={495--507},
  year={2021},
  publisher={IEEE}
}

@article{kumar2023high,
  title={High-fidelity audio compression with improved rvqgan},
  author={Kumar, Rithesh and Seetharaman, Prem and Luebs, Alejandro and Kumar, Ishaan and Kumar, Kundan},
  journal={Advances in Neural Information Processing Systems},
  volume={36},
  pages={27980--27993},
  year={2023}
}

@article{defossez2022high,
  title={High fidelity neural audio compression},
  author={D{\'e}fossez, Alexandre and Copet, Jade and Synnaeve, Gabriel and Adi, Yossi},
  journal={arXiv preprint arXiv:2210.13438},
  year={2022}
}

@article{copet2023simple,
  title={Simple and controllable music generation},
  author={Copet, Jade and Kreuk, Felix and Gat, Itai and Remez, Tal and Kant, David and Synnaeve, Gabriel and Adi, Yossi and D{\'e}fossez, Alexandre},
  journal={Advances in Neural Information Processing Systems},
  volume={36},
  pages={47704--47720},
  year={2023}
}

@article{brown2020language,
  title={Language models are few-shot learners},
  author={Brown, Tom and Mann, Benjamin and Ryder, Nick and Subbiah, Melanie and Kaplan, Jared D and Dhariwal, Prafulla and Neelakantan, Arvind and Shyam, Pranav and Sastry, Girish and Askell, Amanda and others},
  journal={Advances in neural information processing systems},
  volume={33},
  pages={1877--1901},
  year={2020}
}

@article{ho2022classifier,
  title={Classifier-free diffusion guidance},
  author={Ho, Jonathan and Salimans, Tim},
  journal={arXiv preprint arXiv:2207.12598},
  year={2022}
}

@article{zen2019libritts,
  title={Libritts: A corpus derived from librispeech for text-to-speech},
  author={Zen, Heiga and Dang, Viet and Clark, Rob and Zhang, Yu and Weiss, Ron J and Jia, Ye and Chen, Zhifeng and Wu, Yonghui},
  journal={arXiv preprint arXiv:1904.02882},
  year={2019}
}

@article{pratap2020mls,
  title={Mls: A large-scale multilingual dataset for speech research},
  author={Pratap, Vineel and Xu, Qiantong and Sriram, Anuroop and Synnaeve, Gabriel and Collobert, Ronan},
  journal={arXiv preprint arXiv:2012.03411},
  year={2020}
}

@article{anastassiou2024seed,
  title={Seed-tts: A family of high-quality versatile speech generation models},
  author={Anastassiou, Philip and Chen, Jiawei and Chen, Jitong and Chen, Yuanzhe and Chen, Zhuo and Chen, Ziyi and Cong, Jian and Deng, Lelai and Ding, Chuang and Gao, Lu and others},
  journal={arXiv preprint arXiv:2406.02430},
  year={2024}
}

@article{ho2020denoising,
  title={Denoising diffusion probabilistic models},
  author={Ho, Jonathan and Jain, Ajay and Abbeel, Pieter},
  journal={Advances in neural information processing systems},
  volume={33},
  pages={6840--6851},
  year={2020}
}

@inproceedings{sohl2015deep,
  title={Deep unsupervised learning using nonequilibrium thermodynamics},
  author={Sohl-Dickstein, Jascha and Weiss, Eric and Maheswaranathan, Niru and Ganguli, Surya},
  booktitle={International conference on machine learning},
  pages={2256--2265},
  year={2015},
  organization={pmlr}
}

@article{song2020score,
  title={Score-based generative modeling through stochastic differential equations},
  author={Song, Yang and Sohl-Dickstein, Jascha and Kingma, Diederik P and Kumar, Abhishek and Ermon, Stefano and Poole, Ben},
  journal={arXiv preprint arXiv:2011.13456},
  year={2020}
}

@article{kim2020glow,
  title={Glow-tts: A generative flow for text-to-speech via monotonic alignment search},
  author={Kim, Jaehyeon and Kim, Sungwon and Kong, Jungil and Yoon, Sungroh},
  journal={Advances in Neural Information Processing Systems},
  volume={33},
  pages={8067--8077},
  year={2020}
}

@inproceedings{popov2021grad,
  title={Grad-tts: A diffusion probabilistic model for text-to-speech},
  author={Popov, Vadim and Vovk, Ivan and Gogoryan, Vladimir and Sadekova, Tasnima and Kudinov, Mikhail},
  booktitle={International conference on machine learning},
  pages={8599--8608},
  year={2021},
  organization={PMLR}
}

@inproceedings{
shen2024naturalspeech,
title={NaturalSpeech 2: Latent Diffusion Models are Natural and Zero-Shot Speech and Singing Synthesizers},
author={Kai Shen and Zeqian Ju and Xu Tan and Eric Liu and Yichong Leng and Lei He and Tao Qin and sheng zhao and Jiang Bian},
booktitle={The Twelfth International Conference on Learning Representations},
year={2024},
url={https://openreview.net/forum?id=Rc7dAwVL3v}
}

@inproceedings{lavechin2023brouhaha,
  title={Brouhaha: multi-task training for voice activity detection, speech-to-noise ratio, and C50 room acoustics estimation},
  author={Lavechin, Marvin and M{\'e}tais, Marianne and Titeux, Hadrien and Boissonnet, Alodie and Copet, Jade and Rivi{\`e}re, Morgane and Bergelson, Elika and Cristia, Alejandrina and Dupoux, Emmanuel and Bredin, Herv{\'e}},
  booktitle={2023 IEEE Automatic Speech Recognition and Understanding Workshop (ASRU)},
  pages={1--7},
  year={2023},
  organization={IEEE}
}

@inproceedings{shen2018natural,
  title={Natural tts synthesis by conditioning wavenet on mel spectrogram predictions},
  author={Shen, Jonathan and Pang, Ruoming and Weiss, Ron J and Schuster, Mike and Jaitly, Navdeep and Yang, Zongheng and Chen, Zhifeng and Zhang, Yu and Wang, Yuxuan and Skerrv-Ryan, Rj and others},
  booktitle={2018 IEEE international conference on acoustics, speech and signal processing (ICASSP)},
  pages={4779--4783},
  year={2018},
  organization={IEEE}
}

@inproceedings{tokuda2000speech,
  title={Speech parameter generation algorithms for HMM-based speech synthesis},
  author={Tokuda, Keiichi and Yoshimura, Takayoshi and Masuko, Takashi and Kobayashi, Takao and Kitamura, Tadashi},
  booktitle={2000 IEEE international conference on acoustics, speech, and signal processing. Proceedings (Cat. No. 00CH37100)},
  volume={3},
  pages={1315--1318},
  year={2000},
  organization={IEEE}
}

@article{zen2009statistical,
  title={Statistical parametric speech synthesis},
  author={Zen, Heiga and Tokuda, Keiichi and Black, Alan W},
  journal={speech communication},
  volume={51},
  number={11},
  pages={1039--1064},
  year={2009},
  publisher={Elsevier}
}

@article{tokuda2013speech,
  title={Speech synthesis based on hidden Markov models},
  author={Tokuda, Keiichi and Nankaku, Yoshihiko and Toda, Tomoki and Zen, Heiga and Yamagishi, Junichi and Oura, Keiichiro},
  journal={Proceedings of the IEEE},
  volume={101},
  number={5},
  pages={1234--1252},
  year={2013},
  publisher={IEEE}
}

@inproceedings{arik2017deep,
  title={Deep voice: Real-time neural text-to-speech},
  author={Ar{\i}k, Sercan {\"O} and Chrzanowski, Mike and Coates, Adam and Diamos, Gregory and Gibiansky, Andrew and Kang, Yongguo and Li, Xian and Miller, John and Ng, Andrew and Raiman, Jonathan and others},
  booktitle={International conference on machine learning},
  pages={195--204},
  year={2017},
  organization={PMLR}
}

@inproceedings{ping2018deep,
  title={Deep voice 3: 2000-speaker neural text-to-speech},
  author={Ping, Wei and Peng, Kainan and Gibiansky, Andrew and Arik, Sercan O and Kannan, Ajay and Narang, Sharan and Raiman, Jonathan and Miller, John},
  booktitle={proc. ICLR},
  volume={79},
  pages={1094--1099},
  year={2018}
}

@article{gibiansky2017deep,
  title={Deep voice 2: Multi-speaker neural text-to-speech},
  author={Gibiansky, Andrew and Arik, Sercan and Diamos, Gregory and Miller, John and Peng, Kainan and Ping, Wei and Raiman, Jonathan and Zhou, Yanqi},
  journal={Advances in neural information processing systems},
  volume={30},
  year={2017}
}

@article{wang2017tacotron,
  title={Tacotron: A fully end-to-end text-to-speech synthesis model},
  author={Wang, Yuxuan and Skerry-Ryan, RJ and Stanton, Daisy and Wu, Yonghui and Weiss, Ron J and Jaitly, Navdeep and Yang, Zongheng and Xiao, Ying and Chen, Zhifeng and Bengio, Samy and others},
  journal={arXiv preprint arXiv:1703.10135},
  volume={164},
  year={2017},
  publisher={CoRR}
}

@article{sotelo2017char2wav,
  title={Char2wav: End-to-end speech synthesis},
  author={Sotelo, Jose and Mehri, Soroush and Kumar, Kundan and Santos, Joao Felipe and Kastner, Kyle and Courville, Aaron and Bengio, Yoshua},
  year={2017}
}

@article{mehri2016samplernn,
  title={SampleRNN: An unconditional end-to-end neural audio generation model},
  author={Mehri, Soroush and Kumar, Kundan and Gulrajani, Ishaan and Kumar, Rithesh and Jain, Shubham and Sotelo, Jose and Courville, Aaron and Bengio, Yoshua},
  journal={arXiv preprint arXiv:1612.07837},
  year={2016}
}

@article{van2016wavenet,
  title={Wavenet: A generative model for raw audio},
  author={Van Den Oord, Aaron and Dieleman, Sander and Zen, Heiga and Simonyan, Karen and Vinyals, Oriol and Graves, Alex and Kalchbrenner, Nal and Senior, Andrew and Kavukcuoglu, Koray and others},
  journal={arXiv preprint arXiv:1609.03499},
  volume={12},
  year={2016}
}

@inproceedings{panayotov2015librispeech,
  title={Librispeech: an asr corpus based on public domain audio books},
  author={Panayotov, Vassil and Chen, Guoguo and Povey, Daniel and Khudanpur, Sanjeev},
  booktitle={2015 IEEE international conference on acoustics, speech and signal processing (ICASSP)},
  pages={5206--5210},
  year={2015},
  organization={IEEE}
}

@article{coker1976model,
  title={A model of articulatory dynamics and control},
  author={Coker, Cecil H},
  journal={Proceedings of the IEEE},
  volume={64},
  number={4},
  pages={452--460},
  year={1976},
  publisher={IEEE}
}

@misc{speechbrain,
  title={{SpeechBrain}: A General-Purpose Speech Toolkit},
  author={Mirco Ravanelli and Titouan Parcollet and Peter Plantinga and Aku Rouhe and Samuele Cornell and Loren Lugosch and Cem Subakan and Nauman Dawalatabad and Abdelwahab Heba and Jianyuan Zhong and Ju-Chieh Chou and Sung-Lin Yeh and Szu-Wei Fu and Chien-Feng Liao and Elena Rastorgueva and François Grondin and William Aris and Hwidong Na and Yan Gao and Renato De Mori and Yoshua Bengio},
  year={2021},
  eprint={2106.04624},
  archivePrefix={arXiv},
  primaryClass={eess.AS},
  note={arXiv:2106.04624}
}

@misc{ploujnikov2022soundchoice,
      title={SoundChoice: Grapheme-to-Phoneme Models with Semantic Disambiguation}, 
      author={Artem Ploujnikov and Mirco Ravanelli},
      year={2022},
      eprint={2207.13703},
      archivePrefix={arXiv},
      primaryClass={cs.SD}
}

@inproceedings{wang2023wespeaker,
  title={Wespeaker: A research and production oriented speaker embedding learning toolkit},
  author={Wang, Hongji and Liang, Chengdong and Wang, Shuai and Chen, Zhengyang and Zhang, Binbin and Xiang, Xu and Deng, Yanlei and Qian, Yanmin},
  booktitle={ICASSP 2023-2023 IEEE International Conference on Acoustics, Speech and Signal Processing (ICASSP)},
  pages={1--5},
  year={2023},
  organization={IEEE}
}

@article{touvron2023llama,
  title={Llama 2: Open foundation and fine-tuned chat models},
  author={Touvron, Hugo and Martin, Louis and Stone, Kevin and Albert, Peter and Almahairi, Amjad and Babaei, Yasmine and Bashlykov, Nikolay and Batra, Soumya and Bhargava, Prajjwal and Bhosale, Shruti and others},
  journal={arXiv preprint arXiv:2307.09288},
  year={2023}
}

@inproceedings{peebles2023scalable,
  title={Scalable diffusion models with transformers},
  author={Peebles, William and Xie, Saining},
  booktitle={Proceedings of the IEEE/CVF international conference on computer vision},
  pages={4195--4205},
  year={2023}
}

@inproceedings{perez2018film,
  title={Film: Visual reasoning with a general conditioning layer},
  author={Perez, Ethan and Strub, Florian and De Vries, Harm and Dumoulin, Vincent and Courville, Aaron},
  booktitle={Proceedings of the AAAI conference on artificial intelligence},
  volume={32},
  number={1},
  year={2018}
}

@article{loshchilov2017decoupled,
  title={Decoupled weight decay regularization},
  author={Loshchilov, Ilya and Hutter, Frank},
  journal={arXiv preprint arXiv:1711.05101},
  year={2017}
}

\end{document}